\documentstyle[twocolumn,eqsecnum,aps]{revtex}

\def\dx{d^{2}x}

\def\upA{\uparrow}
\def\dnA{\downarrow}

\def\bpmatrix{\left(\begin{array}{c}}
\def\epmatrix{\end{array}\right)}

\def\qh{\text{qh}}
\def\qe{\text{qe}}
\def\pair{\text{pair}}
\def\gap{\text{gap}}
\def\UnitC{E_{\text{C}}^{0}}
\input epsf
\begin{document}
\title{Thermal and Tunneling Pair Creation of Quasiparticles in Quantum Hall 
Systems
}
\author{K. Sasaki and Z. F. Ezawa}
\address{
Department of Physics, Tohoku University, Sendai, 980-8578 Japan
}
\maketitle
\begin{abstract}
We make a semiclassical analysis of thermal pair creations of quasiparticles 
at various filling factors in quantum Hall systems.  It is argued that the gap 
energy is reduced considerably by the Coulomb potential made by impurities.  
It is also shown that a tunneling process becomes important at low temperature 
and at strong magnetic field.  We fit typical experimental data excellently 
based on our semiclassical results of the gap energy.
\end{abstract}

\section{Introduction}

The quantum Hall (QH) effect \cite{DiscoIQHE,DiscoFQHE} has attracted much 
attentions from various points of view.  It is characterized by the appearance 
of Hall plateaux and minima in the longitudinal resistivity.  Observation of a 
zero-resistance state implies the existence of a gap in the excitation 
spectrum leading to the incompressibility of the system.  A Hall plateau 
develops when quasiparticles are pinned by impurities.  Quasiparticles are 
vortices \cite{LaughlinA} and skyrmions \cite{SkyrmQH}.  The aim of this paper is to 
investigate semiclassically the mechanism of thermal creations of 
quasiparticle pairs in the presence of impurities.  It is pointed out that a 
quantum-mechanical tunneling plays an important role in this process.  

Quasiparticles are activated thermally at finite temperature $T$, and 
contribute to the longitudinal current.  It is experimentally known that the 
longitudinal resistivity exhibits a behavior of the Arrhenius type,
\begin{equation}
\rho _{xx} \propto  \exp\biggl(-{\Delta _{\text{gap}}\over 2k_{B}T}\biggr) ,
\label{Arrhenius}
\end{equation}
with $k_{B}$ the Boltzmann constant.  The gap energy $\Delta _{\text{gap}}$ is expected to 
be given by the excitation energy of a pair of quasihole ($\Delta _{\text{qh}}$) and 
quasielectron ($\Delta _{\text{qe}}$).  However, the gap energy experimentally observed 
is much smaller than the theoretical value even if an effect of finite layer 
thickness is taken into account.  Phenomenologically it is well given by
\begin{equation}
\Delta _{\text{gap}}=\Delta _{\text{qh}}+\Delta _{\text{qe}}-\Gamma _{\text{offset}} ,
\label{ActivEnergA}
\end{equation}
with a sample-dependent offset $\Gamma _{\text{offset}}$.  

The offset may be dominated by a Landau-level broadening due to 
impurities \cite{ImpurEffectZR,MacDoLiGiPl}.  They are mainly provided by the 
donors in the bulk situated several hundreds of angstroms away from the 
electron layer.  The Hamiltonian includes the impurity term $H_{\text{imp}}$ 
given by
\begin{equation}
H_{\text{imp}} = e\int \dx \rho (\bbox{x})V_{\text{imp}}(\bbox{x}),
\label{ImpurTerm}
\end{equation}
where $V_{\text{imp}}(\bbox{x})$ is the Coulomb potential made by impurities.  For a 
single impurity it may be approximated by
\begin{equation}
V_{\text{imp}}(\bbox{x}) = \pm {Ze\over 4\pi \varepsilon }{1\over \sqrt {|\bbox{x}|^{2}+d^{2}_{\text{imp}}}},
\label{ImpurPoten}
\end{equation}
where $\pm Ze$ is the impurity charge, $\varepsilon $ is the dielectric constant ($4\pi \varepsilon \simeq 
12.9$), and $d_{\text{imp}}$ is the distance from the layer to the impurity in 
the bulk.

MacDonald et al. \cite{MacDoLiGiPl} derived qualitatively the behavior 
(\ref{ActivEnergA}) by studying an impurity effect on the activation energy of 
magnetorotons \cite{GirvinMcPl} in a perturbation theory, though their predicted 
value for $\Delta _{\text{gap}}$ becomes negative and is physically unacceptable.  
Furthermore, it is not clear how magnetorotons (electrically neutral objects) 
would explain magnetotransport experiments.  See also Ref.\cite{GoldImpurity} for 
a related analysis based on magnetorotons.

We present a simple semiclassical picture for a pair creation of 
quasihole and quasielectron.  Arguing that it occurs to minimize the impurity 
term (\ref{ImpurTerm}), we derive the formula (\ref{ActivEnergA}) with
\begin{equation}
\Gamma _{\text{offset}} \simeq  e^{*}|V_{\text{imp}}(0)|,
\label{EnergGain}
\end{equation}
where a quasiparticle is assumed to be pointlike.  Here, $e^{*}$ is the electric 
charge of quasiholes, $e^{*}=e/m$ at the filling factor $\nu =n/m$ with odd $m$ 
($m=1,3,5,\cdots $).  We also argue that thermal activation is aided by a tunneling 
process at sufficiently low temperature and at strong magnetic field.  The 
Arrhenius formula (\ref{Arrhenius}) is generalized as
\begin{equation}
\rho _{xx} \propto  \exp\biggl(-{\Delta _{\gap}\over 2k_{B}T}\biggr)\biggl[1+e^{-S_{\text{tunnel}}/\hbar }\exp\biggl({A^{*}\over k_{B}T}\biggr)\biggr]^{1/2} .
\label{RhoXXTunne}
\end{equation}
This formula contains two energy scales $\Delta _{\gap}$ and $A^{*}$, and $S_{\text{tunnel}}$ 
is the Euclidean action for the tunneling process.

This paper is composed as follows.
In Section II, we summarize theoretical values of gap energies at various 
filling factors.  We then compare them with typical experimental data based on 
the formula (\ref{ActivEnergA}).
In Section III, we discuss semiclassically the dispersion relation of a 
neutral excitation mode made of a quasihole-quasielectron pair.
In Section IV, analyzing thermal creations of quasiparticle pairs, we derive 
the Arrhenius formula (\ref{Arrhenius}) and the generalized formula (\ref{RhoXXTunne}) 
together with (\ref{ActivEnergA}) and (\ref{EnergGain}).  We show that the generalized 
formula gives an excellent fitting of the resistivity $\rho _{xx}$ for typical data.

\section{Gap Energies}

\begin{figure}[thb]
\epsfbox{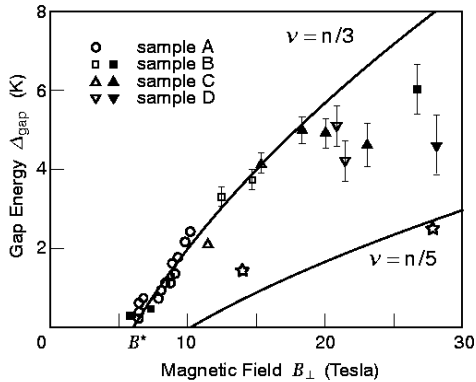}
\caption{
A theoretical result is compared with experimental data for the activation 
energy at $1/3$, $2/3$, $4/3$, $5/3$ and $2/5$, $3/5$ (star symbols).  The 
data are taken from G.S. Boebinger et al. [16].  Theoretical curves are based 
on vortex-excitation formulas (\ref{GapFQHE}).  The impurity potential 
$V_{\text{imp}}(0)$ is taken common for all samples.  See also 
Fig.\ref{BoebTunPS} for the data at $B_{\perp }=8.9$ T and $B_{\perp }=20.9$ T.
}\label{Boeb3PS}
\end{figure}

Vortices are quasiparticles \cite{LaughlinA} in fractional QH states.  They 
have electric charges $\pm e^{*}$ at $\nu =n/m$, where $e^{*}=e/m$.  The excitation energy 
of a vortex pair is solely made of the Coulomb energy,
\begin{equation}
\Delta _{\text{qh}}+\Delta _{\text{qe}} = \alpha _{\pair}^{1/m}\UnitC,
\label{ActivVorte}
\end{equation}
where $\UnitC={e^{2}/4\pi \varepsilon \ell _{B}}$ is the energy unit.  It is expected that
\begin{equation}
\alpha _{\pair}^{1/m} = {1\over m^{2}}\alpha _{\pair} .
\label{AlphaRelat}
\end{equation}
There are several independent estimations on the numerical parameter 
$\alpha _{\pair}^{1/3}$: $\alpha _{\pair}^{1/3}\simeq 0.056$ according to Laughlin \cite{LaughActiv};
$\alpha _{\pair}^{1/3}\simeq 0.053$ according to Chakraborty \cite{ChakrActiv};
$\alpha _{\pair}^{1/3}\simeq 0.094$ according to Morf and Halperin \cite{HalpeActiv};
$\alpha _{\pair}^{1/3}\simeq 0.105$ according to Haldane and Rezayi \cite{HaldaActiv};
$\alpha _{\pair}^{1/3}\simeq 0.106$ according to Girvin, MacDonald and Platzman \cite{GirviActiv};
$\alpha _{\pair}^{1/3}\simeq 0.065$ according to our semiclassical analysis \cite{EzaSasaA}.  Actual 
samples have finite layer widths, which may decrease considerably the Coulomb 
energies \cite{FinitLayerWidth}.  We treat $\alpha _{\pair}^{1/m}$ as a phenomenological 
parameter to analyze experimental data.  As we derive in Section 
\ref{SecThermActiv}, the gap energies at $\nu =n/3$ and $\nu =n/5$ are given by
\begin{mathletters}\label{GapFQHE}
\begin{eqnarray}
\Delta _{\gap}^{1/3}&&= \alpha _{\pair}^{1/3}\UnitC - {e\over 3}|V_{\text{imp}}(0)| , \label{GapFQHEa}\\
\Delta _{\gap}^{1/5}&&= \alpha _{\pair}^{1/5}\UnitC - {e\over 5}|V_{\text{imp}}(0)| . \label{GapFQHEb}
\end{eqnarray}
\end{mathletters}\noindent
We have fitted typical data due to Boebinger et al.\cite{Boeb3} based on these 
formulas in Fig.\ref{Boeb3PS}.  We have used $\alpha _{\pair}^{1/3}=0.50/3^{2}\simeq 0.056$ and 
$\alpha _{\pair}^{1/5}=0.64/5^{2}\simeq 0.026$, where the relation (\ref{AlphaRelat}) holds 
approximately.  We have taken the impurity potential $V_{\text{imp}}(0)$ common 
to all samples, $e|V_{\text{imp}}(0)|=20.4$ K.  It would imply $Z/d_{\text{imp}}\simeq 
1/650 (\AA)$ if the impurity potential (\ref{ImpurPoten}) is assumed.
\epsfxsize=75mm
\epsfxsize=75mm
\begin{figure}[htb]
\epsfbox{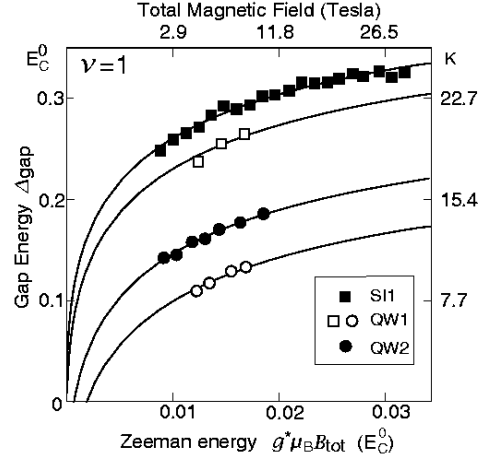}\caption{
A theoretical result is compared with experimental data at $\nu =1$.  The data 
are taken from Schmeller et al. [18].  The theoretical curve is based on the 
skyrmion-excitation formula (\ref{ActivSkyrm}).  The offset $\Gamma _{\text{offset}}$ 
increases as the mobility decreases.  There are two curves for one sample 
(QW1) but with different mobilities.  The mobility changes when electrons are 
pushed against the wall by a bias voltage, as will result in the increase of 
the Coulomb energy $\Gamma _{\text{offset}}$ made by impurities.
}\label{SkyrEneQPS}
\end{figure}

The $\nu =1$ QH state is a QH ferromagnet, where skyrmions \cite{SkyrmQH} are 
excited.  The excitation energy consists of the exchange energy $E_{\text{ex}}$, 
the Coulomb self energy $E_{C}$ and the Zeeman energy $E_{Z}$,  
\begin{eqnarray}
E_{\text{ex}}&&= \sqrt {{\pi \over 32}}E_{C}^{0} , \label{EnergChangSPN}\\
E_{C} &&= {\beta \over 2\kappa }E_{C}^{0} ,  \label{CouloEnerg}\\
E_{Z}&&= {2\widetilde{g}\kappa ^{2}}\ln\biggl({\sqrt {2\pi }\over 32\widetilde{g}}+1\biggr) E_{C}^{0} , \label{NumbeFlip}
\end{eqnarray}
where $\widetilde{g}=g^{*}\mu _{B}B/\UnitC$.  The skyrmion size $\kappa $ is determined to minimize the 
total energy.  The resulting gap energy \cite{EzaSasaA} is
\begin{equation}
\Delta _{\text{gap}}^{1} \simeq  2\biggl(\sqrt {{\pi \over 32}} + {3\beta \over 4\kappa }\biggr)\UnitC - e|V_{\text{imp}}(0)| ,
\label{ActivSkyrm}
\end{equation}
with the skyrmion size
\begin{equation}
\kappa  = {1\over 2}\beta ^{1/3}\biggl\{\widetilde{g}\ln\biggl({\sqrt {2\pi }\over 32\widetilde{g}}+1\biggr)\biggr\}^{-1/3} .
\label{OptimScale}
\end{equation}
The parameter $\beta $ measures the strength of the Coulomb energy, and we have 
$\beta =3\pi ^{2}/64$ for a large skyrmion.  However, an actual skyrmion size is small, 
$\kappa \simeq 1$.  Furthermore, there will be a modification due to a finite thickness of 
the layer \cite{ChakrabortyPietShan}.  We treat $\beta $ as a phenomenological 
parameter.  We have used $\beta =0.24$ to fit typical data \cite{Schmeller} in 
Fig.\ref{SkyrEneQPS}.  The potential $V_{\text{imp}}(0)$ is taken 
phenomenologically as $e|V_{\text{imp}}(0)|\simeq 40 \sim  50$ K.  It would imply 
$Z/d_{\text{imp}}\simeq 2.5/650 (\AA)$ in (\ref{ImpurPoten}).

\begin{figure}[htb]
\epsfbox{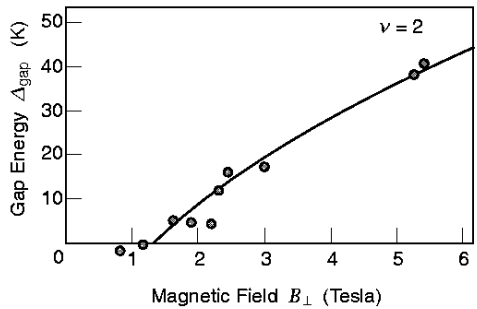}\caption{
A theoretical result is compared with experimental data at $\nu =2$.  The data 
are taken from Usher et al. [20].  The points plotted are those obtained by 
subtracting $\hbar \omega _{c}-g^{*}\mu _{B}B$ from the observed ones.  The theoretical curve is just 
for the Coulomb-energy part and the impurity term in the electron-excitation 
formula (\ref{ActivElect}).  The impurity potential $V_{\text{imp}}(0)$ is taken 
common for all samples.  
}\label{UsherPS}
\end{figure}

Electrons are excited to a higher Landau level and spins are flipped at 
$\nu =2,4,\cdots $.  The gap energy is
\begin{equation}
\Delta _{\gap}^{\nu } = \hbar \omega _{c} + \alpha _{\pair}^{\nu }\UnitC - g^{*}\mu _{B}B - e|V_{\text{imp}}(0)| ,
\label{ActivElect}
\end{equation}
where $\alpha _{\pair}^{\nu }$ is the Coulomb energy associated with the electron-quasihole 
excitation.  It has been estimated that $\alpha _{\pair}^{2}=\sqrt {\pi /8}\simeq 0.63$ by Kallin and 
Halperin \cite{KallinHalperin}.  We have used $\alpha _{\pair}^{2}=0.65$ to fit typical data 
due to Usher et al.\cite{Usher} in Fig.\ref{UsherPS}.  The potential 
$V_{\text{imp}}(0)$ is taken phenomenologically as $e|V_{\text{imp}}(0)|\simeq 37.7 $ K.  
It would imply $Z/d_{\text{imp}}\simeq 2/650 (\AA)$ in (\ref{ImpurPoten}). 
\epsfxsize=75mm

\section{Dispersion Relation}\label{SecDispe}

Thermal fluctuation activates a quasielectron out of the ground state, 
leaving behind a quasihole.  They are created as electrically neutral objects.  
Having charges $\pm e^{*}$ in the magnetic field $B_{\perp }$, with $e^{*}=e/m$ at $\nu =n/m$, 
they feel the Coulomb attractive force as well as the Lorentz force.  We 
examine semiclassically the condition that these two forces are balanced 
\cite{KallinHalperin,YoshiokaRoton}.  Let $V_{\pair}(r)$ be the potential energy of 
the quasiparticle pair with a separation $r$:  The attractive force is 
$\partial V_{\pair}(r)/\partial r$.  The Lorentz force is $e^{*}vB$ when the pair moves parallel to 
the $x$ axis with velocity $v$.  They are balanced when
\begin{equation}
{\partial V_{\pair}(r)\over \partial r} = e^{*}vB .
\label{ForceLoren}
\end{equation}
On the other hand the velocity is given by
\begin{equation}
v={1\over \hbar }{\partial E_{\pair}(\bbox{k})\over \partial k}
\label{VelocDispe}
\end{equation}
in terms of the dispersion relation $E_{\pair}(\bbox{k})$ with $\bbox{k}=(k,0)$.  The total 
energy $E_{\pair}$ is different from the potential energy $V_{\pair}$ by the 
kinetic energy, but it is quenched by the lowest Landau level projection 
\cite{GirvinMcPl}.  Then, we may equate
\begin{equation}
E_{\pair } = V_{\pair}.
\label{RelatEV}
\end{equation}
It follows from (\ref{ForceLoren}), (\ref{VelocDispe}) and (\ref{RelatEV}) that
\begin{equation}
r = mk\ell _{B}^{2} .
\label{PairSepar}
\end{equation}
The dispersion relation $E_{\pair}(\bbox{k})$ of a neutral excitation is obtainable from 
the potential energy $V_{\pair}(r)$ with use of this relation.  This 
semiclassical picture is easily justified by a quantum-mechanical analysis.
\epsfxsize=80mm
\begin{figure}[htb]
\epsfbox{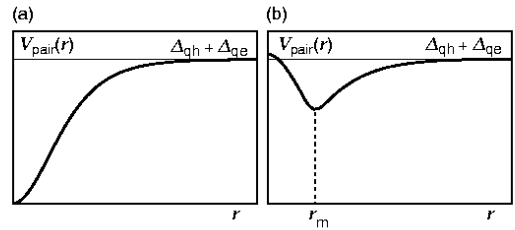}\caption{
The potential energy $V_{\pair}(r)$ of a quasihole-quasielectron pair is 
illustrated.  It may be regarded as the dispersion relation of a neutral 
excitation with use of $r=mk\ell _{B}^{2}$, where the wave vector is given by $(k,0)$.  
(a) The dispersion relation has a gapless mode, which is the case in QH 
ferromagnets.  (b) The dispersion relation may have a minimum point (at 
$r=r_{m}$) describing a magnetoroton, as occurs when a short range repulsive 
interaction acts between a quasihole and a quasielectron.
}\label{PairSkyPS}
\end{figure}

The potential energy $V_{\pair}(r)$ may be approximated by
\begin{equation}
V_{\pair}(r) \simeq  \Delta _{\qh} + \Delta _{\qe} - {e^{*2}\over 4\pi \varepsilon r} ,
\label{ActivEnergB}
\end{equation}
for $r\gg \ell _{B}$.  However, this is a poor approximation for small separation.   
Indeed, according to this formula, $V_{\pair}(r)$ becomes negative for 
sufficiently small $r$.  It is necessary to take into account an overlap of 
quasiparticles.  Quasiparticles are extended objects, vortices and skyrmions, 
described by classical fields.  We place a quasihole at the origin ($\bbox{x}=0$) and 
a quasielectron at the point ($\bbox{x}=\bbox{r}$).  The density modulation is 
$\varrho  _{\pair}(\bbox{x};\bbox{r})=\varrho  _{\text{qh}}(\bbox{x})+\varrho  _{\text{qe}}(\bbox{x}-\bbox{r})$.  The Coulomb energy is
\begin{equation}
V_{\pair}(r) = {1\over 2}{e^{2}\over 4\pi \varepsilon }\int d^{2}xd^{2}x' {\varrho  _{\pair}(\bbox{x};\bbox{r})\varrho  _{\pair}(\bbox{x}';\bbox{r})\over |\bbox{x}-\bbox{x}'|} .
\label{CouloEnergActiv}
\end{equation}
It depends only on the distance $r$ between two quasiparticles provided they 
have cylindrical symmetric configurations.  This is the excitation energy of a 
quasiparticle pair apart from a possible Zeeman energy.  It is reduced to 
(\ref{ActivEnergB}) when two quasiparticles are sufficiently apart.  It is a 
dynamical problem how $\varrho  _{\pair}(\bbox{x};\bbox{r})$ behaves as $\bbox{r}\rightarrow 0$.  We have $V_{\pair}(0)=0$ 
if the quasihole density is precisely cancelled by the quasielectron density, 
$\varrho  _{\text{qh}}(\bbox{x})=-\varrho  _{\text{qe}}(\bbox{x})$, as illustrated in Fig.\ref{PairSkyPS}(a).  It 
implies the existence of a gapless mode in the dispersion relation $E_{\pair}(k)$ 
via the relations (\ref{RelatEV}) and (\ref{PairSepar}).

If the spin degree of freedom is frozen, there exists no cancellation 
since the QH state is incompressible.  Otherwise, a gapless mode which can 
only exists in the density fluctuation would lead to compressibility.  Hence, 
it must be that $\varrho  _{\pair}(\bbox{x};\bbox{r})\not=0$ at $\bbox{r}=0$.  When there exists a short-range 
repulsive interaction between a vortex and an antivortex, the energy 
$E_{\pair}(r)$ may have a minimum describing a magnetoroton at $r=r_{m}\simeq \ell _{B}$ as in 
Fig.\ref{PairSkyPS}(b).  

In QH ferromagnets, on the contrary, the cancellation occurs because 
the dispersion relation contains a gapless mode \cite{CooperFerro}, as illustrated 
in Fig.\ref{PairSkyPS}(a).  A gapless mode develops in the spin fluctuation, 
and hence QH ferromagnets are incompressible in spite of the existence of a 
gapless mode.  By neglecting the Zeeman energy, the perturbative dispersion 
relation is given by \cite{EzaIQCx,Moon},
\begin{equation}
E_{\pair}(\bbox{k}) =  {2\rho _{s}\over \rho _{0}}\bbox{k}^{2},
\label{ZeroModeSpinX}
\end{equation}
as implies
\begin{equation}
V_{\pair}(r) =  {2\rho _{s}\over m^{2}\rho _{0}\ell _{B}^{4}}r^{2}, \quad \quad  \text{at}\quad  r\simeq 0 ,
\end{equation}
where $\rho _{s}$ is the spin stiffness $\rho _{s}={\nu e^{2}/16\sqrt {2\pi }(4\pi \varepsilon )\ell _{B}}$.

\section{Thermal Activation}\label{SecThermActiv}

We study thermal creations of quasiparticle pairs in QH ferromagnets 
with a gapless dispersion relation [Fig.\ref{PairSkyPS}(a)].  We consider two 
cases.  First we analyze a purely thermal process.  We then include a 
tunneling process.  As we shall see, it is obvious that our analysis is 
applicable also to the system where quasiparticles are vortices without 
gapless modes [Fig.\ref{PairSkyPS}(b)].  It is applicable also to certain 
integer QH systems, say at $\nu =2,4,\cdots $, where there are no quasielectrons:  
Here, electrons are activated with quasiholes left behind.  

\subsection{Thermal Process}

At finite temperature $T$, thermal spin fluctuation occurs with the 
rate proportional to the Boltzmann factor $\exp[-E_{\pair}(\bbox{k})/k_{B}T]$ with 
(\ref{ZeroModeSpinX}).  A well-separated quasiparticle pair ($r\rightarrow \infty $) is created with 
rate $\exp[-(\Delta _{\qh}+\Delta _{\qe})/k_{B}T]$, where use was made of (\ref{ActivEnergB}).
\epsfxsize=75mm
\begin{figure}[htb]
\epsfbox{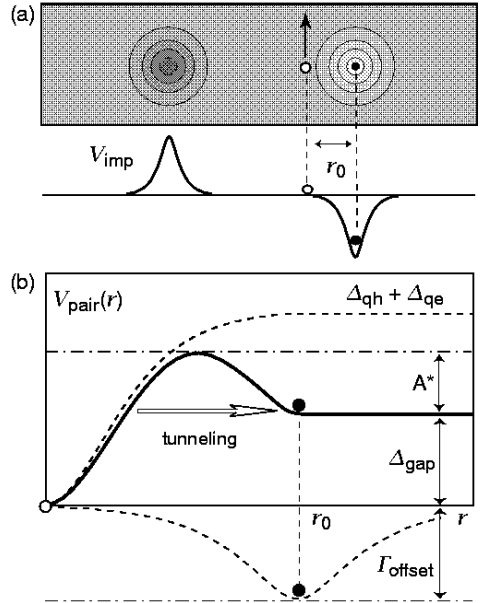}\caption{
(a) An impurity creates a Coulomb potential around it.  It enhances thermal 
activation of a quasiparticle-quasihole pair.  A quasielectron is attracted 
and trapped by the Coulomb potential due to a positive impurity charge, while 
a quasihole is expelled by it.  A quasihole contributes to an Ohmic current.
(b) The creation energy $E_{\pair}(r)$ of a quasiparticle pair is considerably 
reduced by the Coulomb potential due to an impurity charge.  The gap energy of 
one pair is given by 
$\Delta _{\text{gap}}\simeq \Delta _{\qh}+\Delta _{\qe}-\Gamma _{\text{offset}}$, where $\Gamma _{\text{offset}}$ is the energy 
gain.  The effective range of an impurity is denoted by $r_{0}$.
}\label{ActEnePS}
\end{figure}

Thermal activation of quasiparticles is greatly enhanced in the 
presence of impurities bearing electric charges [Fig.\ref{ActEnePS}].  An 
impurity creates a Coulomb potential around it.  For definiteness we assume 
that it has a positive charge.  As we have seen in Section \ref{SecDispe}, 
thermal spin fluctuation is regarded as a creation of a 
quasihole-quasielectron pair.  The pair may be broken near an impurity because 
a quasielectron is attracted by the Coulomb force due to the impurity and a 
quasihole is expelled by it.  The activation energy is given by 
(\ref{ActivEnergA}), where $\Gamma _{\text{offset}}$ is the energy gain (\ref{EnergGain}) when 
the quasiparticle is trapped by a charged impurity [Fig.\ref{ActEnePS}(b)].  
When a quasielectron is trapped by an impurity, only a quasihole moves and 
contributes to an Ohmic current [Fig.\ref{ActEnePS}(a)].

We estimate the number density of quasiparticles in thermal equilibrium 
at temperature $T$.  On one hand, activated from the ground state near an 
impurity, a quasiparticle is transferred to the center of the impurity 
[Fig.\ref{ActEnePS}(b)].  The height of the potential barrier to jump over is 
$A^{*}+\Delta _{\text{gap}}$.  The transition rate is
\begin{equation}
R_{\upA } = c\rho _{0}\exp\biggl(-{A^{*}+\Delta _{\text{gap}}\over k_{B}T}\biggr) ,
\label{UpTherm}
\end{equation}
where $c$ is a constant depending on the density of impurities.  On the other 
hand, recombined with a quasihole, a quasielectron is transferred back to the 
ground state.  The height of the potential barrier to jump over is $A^{*}$.  The 
transition rate is
\begin{equation}
R_{\dnA } = n_{\text{qh}}n_{\text{qe}}\sigma _{\pair}\exp\biggl(-{A^{*}\over k_{B}T}\biggr) ,
\label{DownTherm}
\end{equation}
where $n_{\text{qh}}$ and $n_{\text{qe}}$ are the number densities of quasiholes and 
quasielectrons; $\sigma _{\pair}$ is a certain cross section.  When the system is at 
thermal equilibrium there exists a detailed balance between these two 
transitions, $R_{\upA }=R_{\dnA }$, from which we derive
\begin{equation}
n_{\text{qh}}n_{\text{qe}}={c\rho _{0}\over \sigma _{\pair}}\exp\biggl(-{\Delta _{\text{gap}}\over k_{B}T}\biggr) .
\end{equation}
Since quasiholes and quasiparticles are activated in pairs, we find
\begin{equation}
n_{\text{qh}}=n_{\text{qe}}=n_{0}\exp\biggl(-{\Delta _{\text{gap}}\over 2k_{B}T}\biggr) ,
\end{equation}
at the center of the plateau, where $n_{0}=\sqrt {{c\rho _{0}/\sigma _{\pair}}}$.  The Ohmic current 
is given by the formula (\ref{Arrhenius}) with (\ref{ActivEnergA}) since it is 
proportional to the number density of quasiparticles.  

The QH system is unstable when the gap energy $\Delta _{\text{gap}}$ becomes 
negative.  QH states break down when
\begin{equation}
\Delta _{\text{qh}}+\Delta _{\text{qe}} < \Gamma _{\text{offset}} .
\label{BreakDownB}
\end{equation}
The excitation energy of the pair decreases as the magnetic field decreases.  
The critical magnetic field is derived from (\ref{BreakDownB}),
\begin{equation}
B_{\perp }^{*} = m^{2} {16\pi ^{2}\varepsilon ^{2}\hbar \over e^{3}\alpha _{\pair}^{1/m}}|V_{\text{imp}}(0)|^{2} ,
\end{equation}
for vortex activation (\ref{ActivVorte}) at $\nu =n/m$.  QH states do not exist for 
$B<B_{\perp }^{*}$.  This is consistent with typical data [Fig.\ref{Boeb3PS}].
\epsfxsize=75mm
\begin{figure}[htb]
\epsfbox{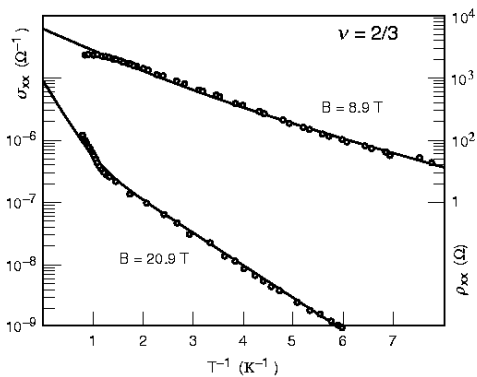}\caption{
Temperature dependence of the minimum of the longitudinal conductance $\sigma _{xx}$ 
and resistance $\rho _{xx}$ at $\nu =2/3$.  The data are taken from G.S. Boebinger et 
al. [16].  Theoretical curves are given by the generalized formula 
(\ref{NumbeTunne}) with vortex excitation (\ref{GapFQHEa}).  See also Fig.\ref{Boeb3PS}.
}\label{BoebTunPS}
\end{figure}

\subsection{Tunneling Process}

We have so far considered a purely thermal process of pair creation.  
However, a tunneling process enhances thermal activation at sufficiently low 
temperature.  When a pair of quasiparticles acquires an energy $\Delta _{\gap}$ 
thermally, it can tunnel across the potential barrier with height $A^{*}$ as in 
Fig.\ref{ActEnePS}.  The transition rate is 
\begin{equation}
R^{\text{tunnel}}_{\upA } = c\rho _{0}e^{-S_{\text{tunnel}}/\hbar }\exp\biggl(-{\Delta _{\text{gap}}\over k_{B}T}\biggr) ,
\label{UpTunne}
\end{equation}
where $S_{\text{tunnel}}$ is the Euclidean action for the tunneling process.  It 
depends on the height $A^{*}$ and the range $r_{0}$.  It is obvious the transition 
rate (\ref{UpTunne}) dominates the rate (\ref{UpTherm}) as $T\rightarrow 0$.  The rate of 
recombination process is still given by (\ref{DownTherm}), because of the plateau 
in the potential for $r>r_{0}$ in Fig.\ref{ActEnePS}.  The detailed balance 
implies
\begin{equation}
R_{\upA } + R_{\upA }^{\text{tunnel}}=R_{\dnA } ,
\end{equation}
with (\ref{UpTherm}), (\ref{DownTherm}) and (\ref{UpTunne}), from which we obtain
\begin{eqnarray}
&& n_{\text{qh}}=n_{\text{qe}} \nonumber\\
&& =n_{0}\exp\biggl(-{\Delta _{\gap}\over 2k_{B}T}\biggr)\biggl[1+e^{-S_{\text{tunnel}}/\hbar }\exp\biggl({A^{*}\over k_{B}T}\biggr)\biggr]^{1/2} .
\label{NumbeTunne}
\end{eqnarray}
This formula contains two energy scales $\Delta _{\gap}$ and $A^{*}$.  We have fitted 
typical data due to Boebinger et al.\cite{Boeb3} in Fig.\ref{BoebTunPS}.  In so 
doing we have determined $\Delta _{\gap}$ by our theoretical formula (\ref{GapFQHEa}), 
$\Delta _{\gap}=\Delta _{\gap}^{1/3}$ with use of $\Gamma _{\text{offset}}={1\over 3}e|V_{\text{imp}}(0)|=6.8$ K.  
The theoretical curve (for $B=8.9$ T) is obtained by using $\Delta _{\gap}\simeq 1.7$ K, $A^{*}\simeq 
0.69$ K and $S_{\text{tunnel}}/\hbar \simeq 2.0$.  The theoretical curve (for $B=20.9$ T) is 
obtained by using $\Delta _{\gap}\simeq 6.1$ K, $A^{*}\simeq 3.7$ K and $S_{\text{tunnel}}/\hbar \simeq 4.0$.  The 
tunneling process makes an important contribution at strong magnetic field 
because $A^{*}$ becomes larger.  They explain quite well the temperature 
dependence of the minimum of the longitudinal resistance $\rho _{xx}$.  

\section{Discussions}

We have analyzed semiclassically a mechanism of thermal and tunneling 
pair creations of quasiparticles in the presence of impurities.  Our formulas 
(\ref{ActivEnergA}) with (\ref{EnergGain}) account for experimental data quite well.  
The impurity effect is summarized into the parameter $Z/d_{\text{imp}}$ in 
(\ref{ImpurPoten}).  We list characteristic features at various filling factors.

(A) Experimental data by Boebinger et at.\cite{Boeb3} at fractional filling 
factors are explained by excitation of vortices with $Z/d_{\text{imp}}\simeq 1/650 
(\AA)$.  Activation energy is rather insensitive to samples.

(B) Experimental data by Usher et at.\cite{Usher} at $\nu =2$ are explained by 
excitation of electrons into higher Landau level with $Z/d_{\text{imp}}\simeq 2/650 
(\AA)$.  Activation energy is rather insensitive to samples.

(C) Experimental data by Schmeller et at.\cite{Schmeller} at $\nu =1$ are 
explained by excitation of skyrmions with $Z/d_{\text{imp}}\simeq 2.5/650 (\AA)$.  
Activation energy is sensitive to sample movilities.

These numbers ($Z=1\sim 3$ and $d_{\text{imp}}\simeq 650 \AA$) are quite reasonable.  
If we take the results seriously, it seems that only skyrmions are sensible to 
sample movilities.  This might be related to the fact that the skyrmion has no 
intrinsic size.  However, we wish to urge caution.  First of all, our 
semiclassical analysis is the first order approximation to the problem, and 
further improvements will be necessary.  For instance, we have assumed that 
quasiparticles are pointlike objects to derive the gap-energy formula 
(\ref{ActivEnergA}) with (\ref{EnergGain}).  It is clear in Fig.\ref{ActEnePS} that the 
formula should be modified when the overlap of quasihole and quasielectron is 
not negligible at their dissociation range $r_{0}$.  Second, experimental data 
are taken from different sources at different dates.  It is necessary to make 
careful experiments by using a single sample to determine the parameter 
$Z/d_{\text{imp}}$ at various filling factors.  We wish to propose such 
experiments.

We have pointed out the importance of tunneling process in thermal 
activation at sufficiently small temperature and at strong magnetic field.  It 
is remarkable that the temperature dependence of the minimum of the 
longitudinal resistance $\rho _{xx}$ is fitted excellently by our formula 
(\ref{RhoXXTunne}) over a wide range of temperature.  This formula is very 
different from any of previously proposed ones \cite{RhoXX}.  QH systems may 
acquire additional interest from the importance of tunneling process.  We 
would like to make a quantitative analysis of this tunneling process in a 
future report.

\end{document}